\documentclass[11pt]{article}
\usepackage{epsfig}
\usepackage{amssymb}
\def\ktwo{{\hat{\kappa}}^2}
\def\kfour{{\hat{\kappa}}^4}
\def\lambrs{{\Lambda}_{RS}}

\def\etal{{\it et al.}}
\def\Journal#1#2#3#4{{#1} {\bf #2}, #3 (#4)}

\def\AJ{\em Astro. J.}
\def\APJ{\em ApJ.}

\def\NPB{{\em Nucl. Phys.} B}
\def\PLB{{\em Phys. Lett.}  B}
\def\PRL{\em Phys. Rev. Lett.}
\def\PRD{{\em Phys. Rev.} D}

\def\be{\begin{equation}}
\def\ee{\end{equation}}
\def\bea{\begin{eqnarray}}
\def\eea{\end{eqnarray}}

\begin{document}
\begin{center}
\Large \bf {Two-Brane Models and BBN\\}
\end{center} 

\begin{center}
{\it Houri Ziaeepour
\\
eMail: {\tt houriz@hotmail.com}}

\end{center}

\begin {abstract}
We obtain a class of solutions for the AdS$_5$ two-brane models by imposing 
the observed value of cosmological constant and Newton coupling constant on 
the visible brane. 
When all terms up to the first order of matter density are included, the 
cosmological evolution on the observable brane depends on the 
equation of state of the matter and consequently when the pressure exists, the 
cosmology of these models deviates from FLRW cosmology. We show that it is 
possible to choose the 
matter equation of state on the hidden brane to neutralize its contribution 
on the cosmological evolution of the visible brane.  We compare the 
prediction of these models for primordial {\it $^4$He} yield with 
observations. In standard BBN with $n_{\nu}^{light} = 3$ this brane model 
is ruled out. If in addition to 
$3$ SM neutrinos there is one light sterile neutrino, this model reconciles 
the observed {\it $^4$He} yield with a high ${\Omega}_b \sim 0.033 h^{-2}$ 
suggested by BOOMERANG and MAXIMA experiments.

\end {abstract}

\section {Introduction and Conclusions}
Since early works on the cosmology of brane models~\cite{bintury}~\cite{sols}
~\cite{statdep} it is well known that they don't have a standard 
cosmology i.e. the 
evolution of Hubble constant on the observable brane depends on the matter 
density in place of its square root. Nonetheless, it has been 
argued~\cite{sols}~\cite{kantres} that 
when the density of matter is much smaller than the absolute value of the 
brane tension, the effect is negligible and the matter part of $H^2$ evolution 
equation can be linearized. For RS-like models this condition is satisfied 
roughly from before BBN to today and should not have observable consequences 
on the 
light elements yield. Moreover, in the case of one-brane models, special 
choice of $\hat T^{55}$ can retrieve the standard FLRW evolution. These 
solutions are related to the stabilization mechanism of the 
branes~\cite{kanti55}. Two-brane models have additional complexities and the 
matter density on the two branes are coupled. This can seriously influence 
the plausibility of these models.\\
In two recent works~\cite {jullien} and ~\cite {kantres} the solutions of 
two-brane models have been investigated. In ~\cite {jullien} only RS 
models~\cite {rs} with one negative tension brane and one positive tension 
brane are considered. They satisfy the well known relation 
${\rho}_{{\Lambda}_L} = -{\rho}_{{\Lambda}_L} = -{\rho}_B/\mu = 
6\mu/\ktwo$ (see below for definitions). In~\cite {kantres}, Kanti \etal solve the evolution equations for a general case. They apply constraints on the 
hidden brane and their approximations lead to the same cosmological behavior 
on both branes.\\
Here we perform the same 
calculation as in~\cite {kantres} with the difference that we consider all 
relevant terms up to first order of the matter densities in the model. We 
show that in this case, even when the matter 
density is much smaller than the brane tension and the higher order terms are 
negligible, the cosmology on the observable brane deviates from FLRW one and 
depends on the matter equation of state on the brane. It is however possible 
to fine tune the equation of state on the hidden brane such that the 
cosmological evolution of branes decouples. By applying the observational 
constraints i.e. having a very small cosmological constant on the observable 
brane and the solution of the hierarchy problem, we find that at 
least for this subset of solutions, the smallness of the cosmological constant 
and warp factor ($N^2$ as defined below) are related. Moreover, the equation 
of state on the hidden brane becomes very close to the pure cosmological 
constant type.\\
The evolution equation however continues to depend on the pressure. We 
investigate the effect of this unconventional cosmology on the 
primordial nucleosynthesis. By comparing the prediction of these two-brane models for 
{\it $^4$He} yield with observation, we show that for a standard particle 
physics 
model with $n_{\nu}^{light} = 3$ they lead to a too small primordial 
{\it $^4$He}. For $n_{\nu}^{light} = 4$ e.g. if there is one strile neutrino, 
this class of brane models are compatible with 
the low {\it $^4$He} observation~\cite{lowhelium} and ${\eta}$ in the 
range predicted by BBN. For high baryon density observed by 
new CMB experiments BOOMERANG~\cite{boomerang} and MAXIMA~\cite{maxima}, it is 
compatible with the whole observationally acceptable range of {\it $^4$He}.

\section {Solutions of Two 3-Brane Models}
\begin{figure}[t]
\begin{center}
\psfig{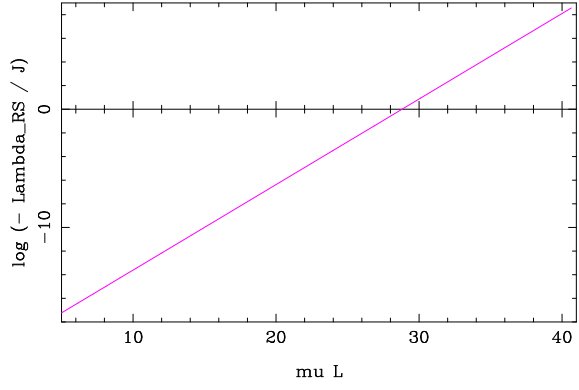}
\caption{Approximative solution of ${\Lambda}_{RS}$ for $M_5 = 10^{15} eV$.
\label {fig:lamb}}
\end{center}
\end{figure}

\begin{figure}[t]
\begin{center}
\psfig{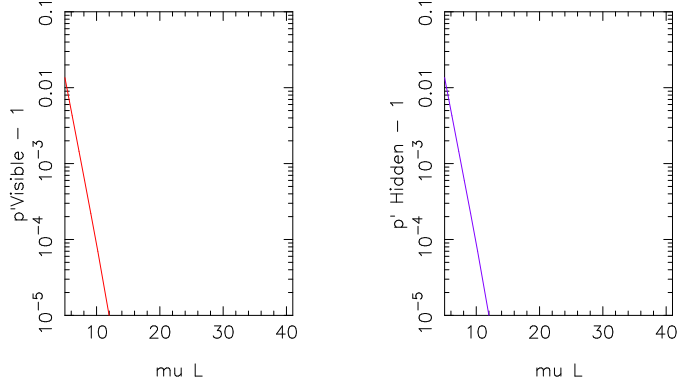}
\caption{${\rho}'_0$ and ${\rho}'_L$ from the approximative solution 
(\ref {approxsol}) for $M_5 = 10^{15} eV$.\label {fig:rhop}}
\end{center}
\end{figure}

The start point of the model is the assumption of one extra dimension. 
Motivated by orbifold compactification of one of space dimensions in string 
theory on ${\bf S_1} / {\bf Z_2}$~\cite{witten}, the positive and negative 
side of the fifth dimension are identified. Considering a homogeneous metric 
for other space dimensions, the metric is defined as:
\be
ds^2 = -n^2 (t, y) dt^2 + a^2 (t, y) \delta_{ij} dx^i dx^j + b^2 (t, y) dy^2. 
\label {metric}
\ee
The orbifoldized dimension is bounded by two 3-branes with coordinate 
separation $L$. If 
$L \rightarrow \infty$, one forgets the brane at infinity (i.e. regulator 
brane) and the brane at $y = 0$ is identified with our 4-dim. Universe. The 
action of this model is defined as:
\be
S = - \int d^4x dy \sqrt {(-\hat {g})} \biggl (\frac {\hat {R}}{2 \ktwo} 
 + {\Lambda}_B + {\Lambda}_0 \delta (y) + {\Lambda}_L \delta (y - L) + 
\hat {\mathcal L}_m \biggr ). \label {action}
\ee
Hatted quantities are in 5-dimensional space. The Einstein equations becomes:
\bea
\hat {G}_{00} & = & 3 \biggl \{ \frac {\dot {a}}{a} \biggl ( 
\frac {\dot {a}}{a} + 
\frac {\dot {b}}{b} \biggr ) - \frac {n^2}{b^2} \biggl [\frac {a''}{a} + 
\frac {a'}{a} \biggl ( \frac {a'}{a} + \frac {b'}{b} \biggr ) \biggr ] 
\biggr \} = \ktwo \hat {T}_{00}, \label {g00} \\
\hat {G}_{ij} & = & \frac {a^2}{b^2} \biggl \{ \frac {a'}{a} \biggl ( 
\frac {a'}{a} + 2 \frac {n'}{n} \biggr ) - \frac {b'}{b} \biggl ( 
\frac {n'}{n} + 2 \frac {a'}{a} \biggr ) + 2 \frac {a''}{a} + \frac {n''}{n} 
\biggr \} \nonumber \\
 & + & \frac {a^2}{n^2} \biggl \{ \frac {\dot {a}}{a} \biggl ( 
- \frac {\dot {a}}{a} + 2 \frac {\dot {n}}{n} \biggr ) - 2 \frac {\ddot {a}}{a} 
+ \frac {\dot {b}}{b} \biggl ( -2 \frac {\dot {a}}{a} + \frac {\dot {n}}{n} 
\biggr ) - \frac {\ddot {b}}{b} \biggr \} = \ktwo \hat {T}_{ii}, 
\label {gii} \\
\hat {G}_{05} & = & 3 \biggl ( \frac {n' \dot {a}}{n a} + \frac {a' 
\dot {b}}{a b} - \frac {\dot {a'}}{a} \biggr ) = 0, \label {g05} \\
\hat {G}_{55} & = & 3 \biggl \{ \frac {a'}{a} \biggl ( \frac {a'}{a} + 
\frac {n'}{n} \biggr ) - \frac {b^2}{n^2} \biggl [\frac {\dot {a}}{a} 
\biggl ( \frac {\dot {a}}{a} - \frac {\dot {n}}{n} + \frac {\ddot {a}}{a} 
\biggr ) \biggr ] \biggr \} = \ktwo \hat {T}_{55}. \label {g55}
\eea
The parameter $\hat {\kappa}^2 = 8 \pi / M^3_{5}$ is the gravity coupling 
constant and 
$\hat {T}_{AB}, \quad A, B = {0, \ldots, 3 \& 5}$ is the energy-momentum 
tensor in 5-dim. space-time. We define $\hat {T}^A_B = \hat {T}^A_{bulk B} + 
\hat {T}^A_{0 B} + \hat {T}^A_{L B}$ as the following:
\bea
\hat {T}^A_{bulk B} & = & diag (-{\rho}_B, P_B, P_B, P_B, \hat {T}^5_5), 
\label {tbulk}\\
\hat {T}^A_{0 B} & = & \frac {\delta (y)}{b} diag (-{\rho}_0, P_0, P_0, 
P_0, 0), \label {t0}\\
\hat {T}^A_{L B} & = & \frac {\delta (y - L)}{b} diag (-{\rho}_L, P_L, P_L, 
P_L, 0). \label {tr}
\eea
${\rho}_i = {\rho}_{m_i} + {\rho}_{{\Lambda}_i}$, $P_i = P_{m_i} - 
{\rho}_{{\Lambda}_i}$, $i$ = $0$ or $L$. It is assumed that they satisfy 
Bianchi identities:
\bea
\dot {{\rho}}_B + 3 \frac {\dot {a}}{a} ({\rho}_B + P_B) + \frac {\dot {b}}{b} 
({\rho}_B + \hat {T}^5_5), \label {rhodot} \\
\hat {T'}^5_5 + 3 \frac {a'}{a} (\hat {T}^5_5 - P_B) + \frac {n'}{n} 
({\rho}_B + \hat {T}^5_5). \label {t55p}
\eea
The restriction of (\ref {rhodot}) to branes includes $\dot {b}$ i.e. the 
expansion of the bulk contributes to 
the density conservation on the brane. This would change the evolution of 
the observable Universe and contradict observations. Therefore, we assume 
that $b$ and the distance between branes has been stabilized~\cite{stabil} at 
a very early time and after inflation they are time independent. In this 
case $b$ can be redefine such that it become a constant and normalized to 
$b = 1$.\\
The discontinuity of $a'$ and $n'$ on the branes leads to the following 
boundary conditions:
\bea
\frac {[a']_0}{a_0 b} = - \frac {\ktwo}{3} {\rho}_0, & \quad \quad & 
\frac {[a']_L}{a_L b} = \frac {\ktwo}{3} {\rho}_L. \label {bai} \\
\frac {[n']_0}{n_0 b} = \frac {\ktwo}{3} (2 {\rho}_0 + 3 P_0), & \quad \quad & 
\frac {[n']_L}{n_L b} = - \frac {\ktwo}{3} (2 {\rho}_L + 3 P_L).
\label {bni}
\eea
where $[A]_x = A_{x^+} - A_{x^-}$. It is easy to verify that conditions 
(\ref {bni}) are satisfied once (\ref {bai}) and energy-momentum conservation 
on the branes (restriction of (\ref {rhodot})) are satisfied.\\
From equation (\ref {g05}):
\be
n (t, y) = \frac {\dot {a} (t, y)}{\alpha (t)}. \label {nsol}
\ee
where $\alpha (t)$ is an arbitrary function of $t$. Choosing $n (t,0) = 1$, 
i.e. synchronous gauge on the $y = 0$ brane, $\alpha (t) = \dot {a}_0 (t)$.\\
We assume that ${\rho}_B$ 
and $P_B$ don't depend on $y$. In this case (\ref {rhodot}) is true only if 
these quantities are also time independent i.e. has the form of a cosmological 
constant. With this energy-momentum tensor, (\ref {g00}) and (\ref {g55}) give 
the evolution equation of $a (t, y)$~\cite {bintury}:
\be
(aa')' - {\alpha}^2(t) = \frac {a^2\ktwo T^0_0}{3}. \label {aevol}
\ee
For ${\rho}_B < 0$, equation (\ref {aevol}) has the following solution: 
\be
a^2 (t,y) = A (t) \cosh (\mu y) + B (t) \sinh (\mu y) + C (t). \quad \quad 
\mu = \sqrt {\frac {2 \ktwo}{3} |{\rho|}_B}. \label {asol} 
\ee
Comparing $y$ independent part of (\ref {asol}) and (\ref {g00}) gives:
\be
C (t) = \frac {3 {\alpha}^2}{\ktwo {\rho}_B}. \label {ct}
\ee
$A$ and $B$ are determined from (\ref {bai}) and (\ref {bni}):
\be
A (t) = a^2_0 - c (t), \quad \quad \quad B (t) = - \frac {\ktwo}{3 \mu} 
{\rho}_0 a^2_0. \label {absol}
\ee
Evaluating (\ref {bai}) for $L$-brane using (\ref {asol}) leads to:
\be
\frac {{\alpha}^2 (t)}{a_0^2} = \frac {{\dot {a}}_0^2}{a_0^2} = 
\frac {\ktwo {\rho}_B}{3}\biggl [\frac {({\rho}'_0 {\rho}'_L + 1) \sinh (\mu L)- 
({\rho}'_0 + {\rho}'_L)\cosh (\mu L)}{{\rho}'_L (1-\cosh (\mu L)) + 
\sinh (\mu L)} \biggr ] \label {adot02}
\ee
For any density $\rho$, ${\rho}' \equiv \frac {\rho}{\lambrs}$, 
$\lambrs \equiv \frac {3 \mu}{\ktwo}$. From (\ref {asol}), (\ref {ct}) and 
(\ref {absol}):
\be
\frac {a_L^2 (t)}{a_0^2 (t)} = \frac {{\rho}'_0 (1-\cosh (\mu L) + 
\sinh (\mu L))}{{\rho}'_L (1-\cosh (\mu L)) + \sinh (\mu L)} \label {ala0}
\ee
By differentiating (\ref {ala0}) and using (\ref {adot02}), the evolution 
equation on the visible brane and relation between warp factors can be 
determined:
\bea
\frac {{\dot {a}}_L^2}{a_L^2} & = & \frac {\ktwo {\rho}_B}{3}\biggl [\frac 
{({\rho}'_0 {\rho}'_L + 1) \sinh (\mu L)- ({\rho}'_0 + {\rho}'_L)\cosh (\mu L)}
{({\rho}'_0 (1-\cosh (\mu L)) + \sinh (\mu L))^2} \biggr ] \nonumber \\
 & & \biggl [\frac 
{(2 {\rho}'_{{\Lambda}_0} - {\rho}'_{m_0} - 3 P'_{m_0})(1-\cosh (\mu L)) + 
2 \sinh (\mu L)}{(2 {\rho}'_{{\Lambda}_L} - {\rho}'_{m_L} - 3 P'_{m_L})
(1-\cosh (\mu L)) + 2 \sinh (\mu L)}\biggr ]^2 ({\rho}'_L (1-\cosh (\mu L)) + 
\sinh (\mu L)) \nonumber \\ 
 & & \label {adotl2} \\
\frac {n_L^2}{n_0^2} & = & \frac {a^2_0}{a^2_L} \biggl [\frac 
{(2 {\rho}'_{{\Lambda}_0} - {\rho}'_{m_0} - 3 P'_{m_0})(1-\cosh (\mu L)) + 
2 \sinh (\mu L)}{(2 {\rho}'_{{\Lambda}_L} - {\rho}'_{m_L} - 3 P'_{m_L})
(1-\cosh (\mu L)) + 2 \sinh (\mu L)}\biggr ]^2 \label {nln0}
\eea
With supplementary assumption ${\rho}'_{{\Lambda}_i} \gg {\rho}'_{m_i}$, 
after expansion to first order of matter density, the evolution equation 
(\ref {adotl2}) becomes:
\bea
\frac {{\dot {a}}_L^2}{a_L^2} & = & \frac {\ktwo {\rho}_B {\mathcal A}}{3 
{\mathcal C}} \biggl [1 + \biggl (\frac {{\rho}'_{{\Lambda}_0} \sinh (\mu L) - 
\cosh (\mu L)}{{\mathcal A}} + \frac {2 (1 - \cosh (\mu L))}
{{\mathcal C}} \biggr ) {\rho}'_{m_L} + \nonumber \\
 & & \biggl (\frac {{\rho}'_{{\Lambda}_L} \sinh (\mu L) - \cosh (\mu L)}
{{\mathcal A}} - \frac {3 (1 - \cosh (\mu L))}
{{\mathcal B}} \biggr ) {\rho}'_{m_0} + \nonumber \\
 & & \frac {3 (1 - \cosh (\mu L))}{{\mathcal C}} P'_{m_L} - 
\frac {3 (1 - \cosh (\mu L))}{{\mathcal B}} P'_{m_0} + 
{\mathcal O}({\rho'}_m^2) \biggr ] \label {adotl2obs}\\
{\mathcal A} & \equiv & ({\rho}'_{{\Lambda}_0} {\rho}'_{{\Lambda}_L} + 1) 
\sinh (\mu L) - ({\rho}'_{{\Lambda}_0} + {\rho}'_{{\Lambda}_L})\cosh (\mu L) 
\label {amath} \\
{\mathcal B} & \equiv & {\rho}'_{{\Lambda}_0} (1-\cosh (\mu L)) + \sinh (\mu L)
\label {bmath} \\
{\mathcal C} & \equiv & {\rho}'_{{\Lambda}_L} (1-\cosh (\mu L)) + 
\sinh (\mu L)\label {cmath}
\eea
As noticed in 
~\cite{jullien} and ~\cite{kantres}, the evolution equation on visible brane 
depends on the matter on both branes. Here we see that considering the full 
expansion of (\ref {adotl2}) to first order, not only the evolution depends on 
the matter on both branes but also on their equation of state even at late 
time (In ~\cite{statdep} the same dependence has been obtained for one brane 
models before stabilization). It is in strict 
conflict with the evolution of FLRW metric which only depends on the matter 
density. It is easy to verify that for large $\mu L$, the amplitudes of 
density and pressure terms are comparable and it is not possible to neglect 
the pressure term. This behavior has important consequences for 
nucleosynthesis in the early universe. We address this issue in the next 
section.\\
Equation (\ref {adotl2obs}) has also another interesting consequence. It is 
possible to fine-tune the equation of state of the matter on the hidden brane 
such that it decouples from the cosmological evolution of the visible brane. 
Assuming $P_i = w_i {\rho}_i$, $i = 0$ or $L$ as the equation of state for matter 
and ${\gamma}_0 \equiv 1 + w_0$ and ${\gamma}'_L \equiv \frac{2}{3} + w_L$, 
the value of ${\gamma}_0$ which eliminates the contribution of the matter on 
the hidden brane from (\ref {adotl2obs}) is:
\be
{\gamma}_0 = \frac {{\mathcal B} ({\rho}'_{{\Lambda}_L} \sinh (\mu L) - 
\cosh (\mu L))}{3 {\mathcal A} (1-\cosh (\mu L))} \label {gamm0}
\ee
Our numerical 
calculation shows that for the interesting range of the only 
parameter of the model i.e. $5 < \mu L < 50$, the value of $w_0$ from 
(\ref {gamm0}) is very close to $-1$. This means that if this model corresponds to real 
universe, matter can be absent from the hidden brane or it can be a scalar 
field with a quintessential behavior. It would be interesting to see if 
stabilization models can predict such partition of matter between branes.\\
In the next step we use the linearized equation (\ref{adotl2obs}) to identify 
observable quantities like cosmological constant and Newton coupling constant. 
The visible brane must have an evolution equation similar to FLRW cosmology. 
We parameterize (\ref{adotl2obs}) as the following:
\be
\frac {{\dot {a}}_L^2}{a_L^2} = \frac {8\pi G}{3} (\alpha {\rho}_{m_{obs}} + 
{\rho}_{{\Lambda}_{obs}} + {\mathcal O}({\rho}_m^2)). \label {aobs}
\ee
Quantities ${\rho}_{m_{obs}}$ and ${\rho}_{{\Lambda}_{obs}}$ are 
respectively observed matter density of the Universe and observed 
cosmological constant; $\alpha$ is a dimensionless quantity. For FLRW 
cosmology $\alpha = 1$. For brane models in general $\alpha$ can 
depend on $\mu L$ and the equation of state.\\
In ~\cite{kantres} constraints are imposed on the hidden brane. In fact in 
that work, the square term in (\ref {adotl2}) is neglected and the evolution 
equations on both branes have the same form and there is no 
difference on which brane constraints are imposed. Here we apply them on the 
visible brane. When (\ref {gamm0}) is satisfied, the value of three unknown 
parameters ${\Lambda}_{RS}$, ${\rho}'_{{\Lambda}_0}$ and 
${\rho}'_{{\Lambda}_L}$ can be fixed by comparing (\ref {aobs}) and 
(\ref {adotl2obs}) and an additional condition on the warp 
factor~\cite{kantres}~\cite{nima}:
\bea
{\rho}_{{\Lambda}_{obs}} & = & \frac {\ktwo {\rho}_B}{8 \pi G} \biggl [ \frac 
{{\mathcal A}}{{\mathcal C}}\biggr ] \label {grho} \\
8 \pi G & = & \frac {\ktwo {\rho}_B}{{\Lambda}_{RS}} \biggl [ \frac 
{({\rho}'_{{\Lambda}_0} \sinh (\mu L) - \cosh (\mu L)) {\mathcal C} + 
2 (1 - \cosh (\mu L)) {\mathcal A}}{{\mathcal C}^2} 
\biggr ] \label {pig} \\
N^2 & \equiv & \frac {M^2_5}{M^2_{pl}} = \frac {n^2_L}{n^2_0} = 
\frac {{\mathcal B}}{{\mathcal C}} \label {n2}
\eea
Note that only the value of $\alpha G$ can be determined from (\ref {aobs}) 
and (\ref {adotl2obs}). Here we define $G$ such that:
\be
\alpha = 1 + \beta w_L \label {alphdef} 
\ee
and $\beta$ depends only on $\mu L$ and not on the matter density:
\be
\beta = \frac {3 {\mathcal AN^2} (1 - \cosh (\mu L))}{ 
{\mathcal B}\biggl ({\rho}'_{{\Lambda}_L} \sinh (\mu L) - \cosh (\mu L) \biggr ) - 3 {\mathcal A} \biggl (1 - \cosh (\mu L) \biggr )} \label {betdef}
\ee
The equation (\ref {n2}) is 
obtained after neglecting time dependent terms. 
By applying the same procedure on the quantities of the hidden brane, one finds 
that:
\be
G' = \frac {G}{N^2}\biggl (1 - \frac {3 {\gamma}' (1 - 
\cosh (\mu L)) {\rho}'_{{\Lambda}_{obs}}}{{\mathcal C}^2}\biggr ) \label {ghidden}
\ee
where $G'$ is the 
Newton coupling constant on the hidden brane. According to (\ref {ghidden}) 
the gravitational coupling on the hidden brane is much stronger than on the 
observable one.\\
From equations (\ref {grho}) to (\ref {n2}) one obtains a $5^{th}$ order 
equation for ${\Lambda}_{RS}$. Due to presence of very 
large and very small parameters (respectively $\cosh (\mu L)$, $N^2$ and 
${\rho}_{obs}/ J$ (see (\ref{approxsol}) below for the definition of J)), in 
this equation, one must be very careful about 
approximations because the combination of large and small parameters can 
lead to quantities which are not negligible. For $\cosh (\mu L) \gg 1$ and 
$N^2 \cosh (\mu L) \ll 1$, a simple analytical solution for ${\Lambda}_{RS}$ 
can be found:
\be
{\Lambda}_{RS} = J \biggl (\frac {3 {\gamma}'_L {\rho}^2_{{\Lambda}_{obs}} 
{\cosh}^2 (\mu L)}{N^2 J^2} \biggr )^{\frac {1}{3}} \cosh (\mu L) \quad \quad 
\quad J \equiv \frac {48 \pi G}{\kfour} \label {approxsol}
\ee
It is valid only for $M_5 \gtrsim 10^{14} eV$. Figs. \ref {fig:lamb} and 
\ref {fig:rhop} show 
${\Lambda}_{RS}$, ${\rho}'_{{\Lambda}_0}$, ${\rho}'_{{\Lambda}_L}$ as a 
function of $\mu L$. It is interesting to note that in this approximative 
solution, the smallness of the observed cosmological constant and $N^2$ are 
related. In fact according to this solution, the value of 
${\rho}^2_{{\Lambda}_{obs}}$ and $N^2$ must be roughly of the same order to 
not have too small or too large ${\Lambda}_{RS}$ (for fixed $J$ and 
$\mu L$). It is also evident that in this approximation an exactly null 
cosmological constant is not acceptable because ${\Lambda}_{RS}$ would be 
zero too.\\
Another aspect of this solution is the positiveness of tension on both 
branes. 
Using (\ref {grho}) to (\ref {n2}), it is not difficult to see that when 
$\mu L$ is large, tensions are both very close to ${\Lambda}_{RS}$ (this has 
been also observed in ~\cite{kantres}). A difference of order 
$\cosh^{-1} (\mu L)$ between 
normalized tensions assures the small warp factor Eq. (\ref {n2}).\\

\section {Primordial Nucleosynthesis}
\begin{figure}[t]
\begin{center}
\psfig{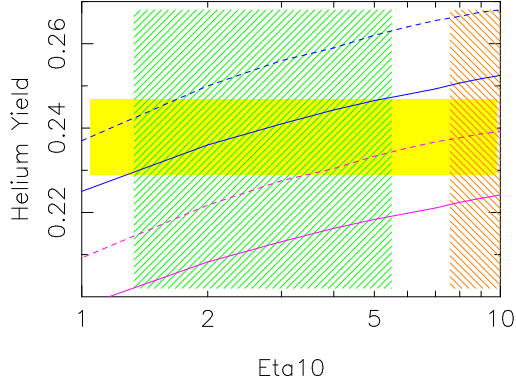}
\caption{Primordial Helium yield: FLRW (blue), model explained in Spec. $2$ 
(magenta). Full and dashed curves correspond respectively to $n^{light}_\nu = 3$ and $n^{light}_\nu = 4$. The hashed regions show a conservative 
range for the observed value of $Y_p$ (yellow) and corresponding ${\eta}_{10}$ according to BBN (light green) and new value of ${\eta}_{10}$ from CMB 
anisotropy observation (red).
\label {fig:yp}}
\end{center}
\end{figure}

In the previous section we have seen that when the matter pressure is not 
negligible, the cosmology of two-brane models deviates from the FLRW 
cosmology even if the higher order terms are negligible. It is therefore 
necessary to determine the prediction of this class of brane models for the 
Big Bang Nucleosynthesis and the yield of the light elements.\\
In a brane universe with cold and hot matter on the visible brane, the 
evolution equation of the visible brane (\ref {aobs}) can be written as:
\be
H^2 \equiv \frac {{\dot {a}}_L^2}{a_L^2} = \frac {8\pi G}{3} ({\alpha}_{hot} 
{\rho}_{hot} + {\rho}_{cold} + {\rho}_{{\Lambda}_{obs}} + 
{\mathcal O}({\rho}_m^2)). \label {coldhot}
\ee
At the time of nucleosynthesis the contribution of higher order terms, cold 
component and cosmological constant are negligible and:
\be
H^2 = \frac {8\pi G}{3}{\alpha}_{hot} {\Omega}_{hot} (1 + z)^4 \label {hot}
\ee
where $z$ is the redshift. Equation (\ref {hot}) has the same form as FLRW 
cosmology with an effective mass of ${\alpha}_{hot} {\Omega}_{hot}$.\\
The relation between primordial yield of light elements depends on the 
temperature of the $p-n$ plasma when neutrinos decouple from weak interaction 
$p + e \leftrightarrows n + \nu$ (see e.g. ~\cite {earlyuniv}). In the 
unconventional cosmology of (\ref {hot}):
\be
\frac {{\Gamma}_{p e \leftrightarrows n \nu}}{H} \approx \frac {1}{{\alpha}^
{\frac {1}{2}}} \biggl (\frac {T}{0.8 MeV } \biggr )^3  \label {dectemp}
\ee
Or in another word $T_{freez-out, brane} \equiv T_F \sim {\alpha}^{\frac {1}
{6}} 0.8 MeV$. 
With this new freeze-out temperature, the neutron-to-proton ratio becomes:
\be
\biggl ( \frac {n}{p} \biggr )_{freeze-out, brane} = \exp \biggl ( -\frac {Q}
{T_F}\biggr ) = \biggl ( \frac {n}{p} \biggr )^{{\alpha}^{-\frac {1}{6}}}_{freeze-out, FLRW}
\ee
When $\mu L \gg 1$ and ${\rho}_0 \approx {\rho}_L$, using (\ref {alphdef}) and 
(\ref {betdef}), $\alpha \approx \frac {2}{3}$. Assuming
\be
\biggl ( \frac {n}{p} \biggr )_{freeze-out, FLRW} \approx \frac {1}{7} = 0.143
\ee
results to:
\be
\biggl ( \frac {n}{p} \biggr )_{freeze-out, brane} \approx 0.125
\ee
i.e. the prediction of the two-brane model studied in the previous section 
is $\sim 12\%$ less than standard cosmology. Fig.\ref {fig:yp} shows 
the {\it $^4$He} yield $Y_p$ as a function of $\eta \equiv n_b/n_\gamma$ 
for FLRW and for the two-brane models and compares them with observations. 
It is evident that for $n^{light}_\nu = 3$ the brane model is ruled out for 
all reasonable values of $\eta$. However, the observation of neutrino 
oscillation by 
Super-Kamiokand experiment and others~\cite{superkam} joint with the 
results of LSAND experiment~\cite {lsand} strongly suggests the 
existence of a sterile neutrino mixing 
with the SM neutrinos. In this case the number of light neutrinos is larger 
than $3$. Fig.\ref {fig:yp} shows also the {\it $^4$He} yield for 
$n^{light}_\nu = 4$ for FLRW and the two-brane models\footnote {The 
effective number of neutrinos must be somehow smaller than $4$ due to the 
oscillation. An initial lepton asymmetry also has the same 
effect~\cite{bbnosc}.}. The 
latter is compatible with the observation specially if ${\Omega}_b$ or 
equivalently $\eta$ is as large as what is suggested by BOOMERANG~\cite{boomerang} and 
MAXIMA~\cite{maxima} experiments. In fact the SBBN has difficulties to 
reconcile ${\Omega}_b$ ($\eta$) obtained from these experiments 
with the independent measurement of {\it $^4$He}~\cite{lowhelium}~\cite
{highhelium} and deuterium~\cite{deut} yields. It is possible to reconcile 
{\it $^4$He} and CMB observations in the standard cosmology if there is a 
sterile neutrino and an initial lepton asymmetry~\cite{bbnosc}. However, to 
have an effective number of neutrinos less than $3$, the sterile neutrino must be 
the lightest one and the mass difference between ${\nu}_s$ and ${\nu}_e$ must 
be $\delta m^2 \lesssim 1 eV^2$~\cite{bbnosc}. Two-brane models by contrast are less 
sensitive to the parameter space of neutrinos and don't need an initial lepton asymmetry.\\
None of FLRW or brane models however can 
reconcile the observed value of $^2D$ if ${\Omega}_b \gtrsim 0.03 h^{-2}$. An 
entropy increase after BBN has been suggested to reconcile two 
observations~\cite{entrop}. In this case, according to Fig.\ref {fig:yp} the 
brane model would be only compatible with a low {\it $^4$He} yield if $\eta$ has a value in 
the range predicted by SBBN.\\
Conclusions of this section are not very sensitive to the details of the 
two-brane model. The 
asymptotic value of $\beta$ is valid for a large range of parameters $M_5$ 
and $\mu L$. We have restricted the analysis to the special model with 
decoupled branes. In the general case the conclusion depends on the density 
and pressure on the hidden brane which are not directly observable.

\end {document}